\documentclass[twocolumn,floatfix,showpacs,prb]{revtex4}
\usepackage{graphicx}
\usepackage{amsmath}
\usepackage{bm}
\begin{document}

\title{Pseudo-diffusive conduction at the Dirac point of a normal-superconductor junction in graphene}
\author{A. R. Akhmerov}
\affiliation{Instituut-Lorentz, Universiteit Leiden, P.O. Box 9506, 2300 RA Leiden, The Netherlands}
\author{C. W. J. Beenakker}
\affiliation{Instituut-Lorentz, Universiteit Leiden, P.O. Box 9506, 2300 RA Leiden, The Netherlands}
\date{October 2006}
\begin{abstract}
A ballistic strip of graphene (width $W\gg$ length $L$) connecting two normal metal contacts is known to have a minimum conductivity of $4e^{2}/\pi h$ at the Dirac point of charge neutrality. We calculate what happens if one of the two contacts becomes superconducting. While the ballistic conductance away from the Dirac point is increased by Andreev reflection at the normal-superconductor (NS) interface, we find that the minimum conductivity stays the same. This is explained as a manifestation of pseudo-diffusive conduction at the Dirac point. As a generalization of our results for a ballistic system, we provide a relation between the conductance $G_{\rm NS}$ of an arbitrarily disordered normal-superconductor junction in graphene and its value $G_{\rm N}$ when both contacts are in the normal state.
\end{abstract}
\pacs{74.45.+c, 73.23.-b, 74.50.+r, 74.78.Na}
\maketitle

\section{Introduction}
\label{intro}

The effect of a superconducting contact on the conductance of a metal is qualitatively different if the electrons propagate ballistically or diffusively through the metal: A superconducting contact doubles the conductance of a ballistic metal, while the conductance of a diffusive metal remains the same. \cite{reviews} The conductance doubling in a ballistic metal happens because an electron incident on the superconductor is Andreev reflected as a hole of opposite charge, hence doubling the current. The effective length $L$ of the conductor is doubled as well, because the hole has to make its way back, and this effect cancels the conductance doubling in the case of a diffusive metal.

These two competing effects are summarized by the approximate relation
\begin{equation}
G_{\rm NS}(L)\approx 2G_{\rm N}(2L),\label{simple}
\end{equation}
between the conductance $G_{\rm NS}$ of a normal-metal--superconductor junction and its normal-state value $G_{\rm N}$. As derived in Ref.\ \onlinecite{Bee92} from a generalized Landauer formula, the relation (\ref{simple}) holds if quantum corrections of order $e^{2}/h$ can be neglected. It also requires an ideal NS interface, without a mismatch of Fermi wave lengths between N and S.

A novel transport regime called ``pseudo-diffusive'' appears in an impurity-free carbon monolayer (= graphene) at zero carrier concentration (= Dirac point). Because of the vanishing density of states the transmission through a strip of undoped graphene (width $W$, length $L$) occurs entirely via evanescent modes. For a short and wide strip there is a large number $N_{\rm eff}=W/L\gg 1$ of evanescent modes with transmission probability of order unity (open channels). Several recent investigations have found that the transport properties of undoped ballistic graphene are similar to those of a diffusive metal with the same number $N_{\rm eff}$ of open channels: Both systems have the same conductance\cite{Kat06,Two06} $G\simeq N_{\rm eff}e^{2}/h$ and the same shot noise power\cite{Two06} (Fano factor $1/3$). In a Josephson junction both systems have the same current-phase relationship, critical current, and current-voltage characteristic.\cite{Tit06,Cue06} This correspondence between evanescent modes in graphene and diffusion modes in a disordered metal is not limited to a carbon monolayer, but applies to a bilayer as well.\cite{Sny06} The effect of disorder on the evanescent modes has also been studied.\cite{Ver06}

In this paper we present one more manifestation of pseudo-diffusion in graphene, by calculating the ratio $G_{\rm NS}/G_{\rm N}$. Far from the Dirac point, at high carrier concentrations, this ratio is between 1 and 2 --- as expected for a ballistic metal with a Fermi wave length mismatch. At the Dirac point, however, the ratio approaches unity --- as in a diffusive metal, but without any scattering by impurities or lattice defects.

The outline of this paper is as follows. In Sec.\ \ref{smatrix} we describe the scattering problem of a lightly doped graphene strip between two heavily doped metal contacts. In Sec.\ \ref{conductance} we calculate the conductance $G_{\rm NS}$ of the NS junction from the probability of Andreev reflection, as a function of the Fermi energy in the graphene strip. We compare with the known results\cite{Kat06,Two06} for $G_{\rm N}$ in Sec.\ \ref{compare}, demonstrating the equality $G_{\rm NS}=G_{\rm N}$ at the Dirac point. This explicit calculation is for a ballistic conductor. In the Appendix we derive a more general relation between $G_{\rm NS}$ and $G_{\rm N}$, valid also for a disordered conductor.

\section{Scattering matrix}
\label{smatrix}

\begin{figure}[tb]
\centerline{\resizebox{0.9\linewidth}{!}{\includegraphics{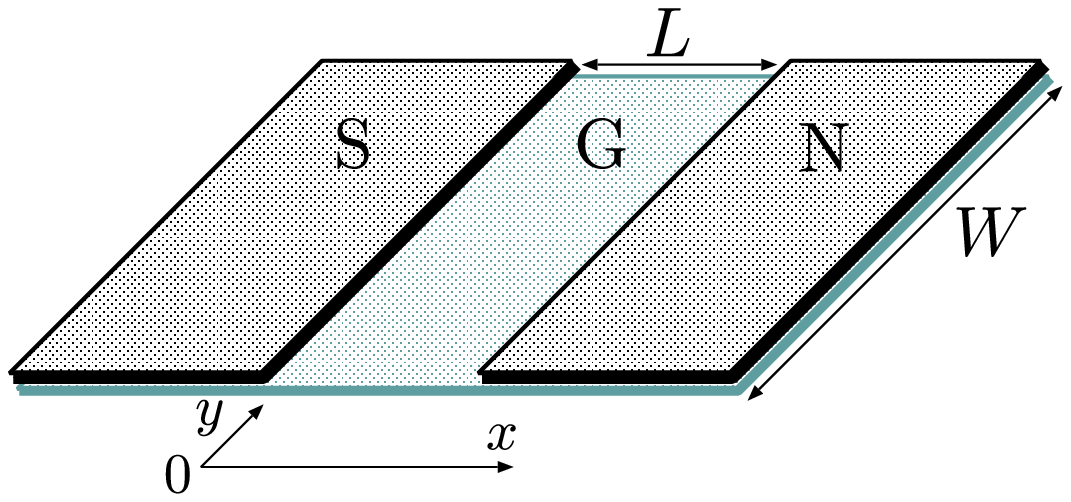}}}
\bigskip
\centerline{\resizebox{0.9\linewidth}{!}{\includegraphics{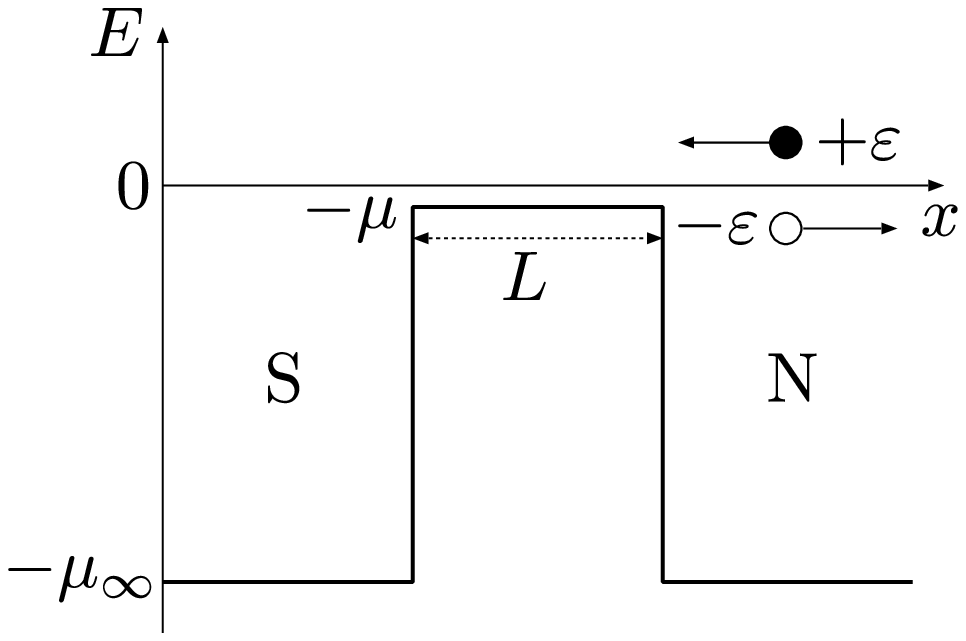}}}
\caption{\label{fig_NGS}
Top panel: geometry of a graphene sheet (G) with one normal (N) and one superconducting contact (S). Bottom panel: Spatial dependence of the Dirac point (solid line). The Dirac point is far below the Fermi level (at zero energy) in the heavily doped contact regions $x<0$ and $x>L$. In the lightly doped central region $0<x<L$ the Dirac point is much closer to the Fermi level. An electron incident from the normal contact (N) above the Fermi level (filled circle) is Andreev reflected by the superconducting contact (S) as a hole below the Fermi level (open circle). In the limit $\mu,\varepsilon\rightarrow 0$ the electrons and holes are transmitted through the central region via evanescent modes.
}
\end{figure}

We consider a graphene sheet having width $W$ in the $y$-direction, with regions $x<0$ (S region) and $x>L$ (N region) covered by superconducting and normal electrodes, respectively. The central region $0<x<L$ is lightly doped, with Fermi energy $\mu$ relative to the Dirac point. The contact regions $x<0$ and $x>L$ are heavily doped, with Fermi energy $\mu_{\infty}$ much larger than both $\mu$ and the superconducting gap $\Delta_{0}$. An electron incident on the superconductor at an energy $\varepsilon$ just above the Fermi level is Andreev reflected as a hole at an energy $\varepsilon$ below the Fermi level. (See Fig.\ \ref{fig_NGS}.) The reflection occurs predominantly via propagating modes for $\mu\gg\hbar v/\min(W,L)$ (with $v$ the energy-independent carrier velocity in graphene). That regime was studied in Refs.\ \onlinecite{Bee06,Sen06}. In the limit $\mu\rightarrow 0$ at fixed $W,L$ only evanescent modes remain. This is the regime studied here.

We calculate the scattering matrix of the NS junction using the Dirac-Bogoliubov-de-Gennes equation,\cite{Bee06}
\begin{equation}
\begin{pmatrix}
v \bm{p}\cdot\bm{\sigma}+U& \Delta \\
\Delta^{\ast} & -v \bm{p}\cdot\bm{\sigma}-U
\end{pmatrix}\Psi=\varepsilon\Psi,\label{DBdG}
\end{equation}
where $\bm{p}=-i\hbar(\partial/\partial x,\partial/\partial y)$ is the momentum operator and $\bm{\sigma}=(\sigma_{x},\sigma_{y})$ is the isospin operator. The excitation energy $\varepsilon>0$ is measured relative to the Fermi level (set at zero). The electrostatic potential $U$ and pair potential $\Delta$ have step function profiles,
\begin{eqnarray}
U(x)&=&\left\{\begin{array}{ll}
-\mu_{\infty}&{\rm if}\;\;x<0\;\;{\rm or}\;\;x>L,\\
-\mu&{\rm if}\;\;0<x<L,
\end{array}
\right.\label{Udef}\\
\Delta(x)&=&\left\{\begin{array}{ll}
\Delta_{0}&{\rm if}\;\;x<0,\\
0&{\rm if}\;\;x>0.
\end{array}
\right.\label{Deltadef}
\end{eqnarray}
Because there is only a single superconductor, we can take $\Delta_{0}$ real without loss of generality.

The Fermi wave length $\lambda'_{F}=hv/\mu_{\infty}$ in the contact regions is sent to zero by taking the limit $\mu_{\infty}\rightarrow\infty$. The infinitely large mismatch with the Fermi wave length $\lambda_{F}=hv/\mu$ in the central region would fully suppress Andreev reflection of nonrelativistic electrons.\cite{And64} It is a special property of the relativistic wave equation (\ref{DBdG}) that a nonzero  Andreev reflection probability remains regardless of the Fermi wave length mismatch.\cite{Bee06}

For aspect ratios $W\gg L$ the boundary conditions in the $y$-direction are irrelevant. We take periodic boundary conditions for simplicity. Different wave vectors $q_{n}=2\pi n/W$ (with $n=0,\pm 1,\pm 2,\ldots$) in the $y$-direction are not coupled, so we can consider each transverse mode separately. In the normal regions there are four eigenstates of Eq.\ (\ref{DBdG}) for each $n$ and $\varepsilon$, corresponding to electrons and holes propagating in either $+x$ or $-x$ direction. In the superconducting region there are two eigenstates that decay for $x\rightarrow -\infty$. These eigenstates are given in Ref.\ \onlinecite{Bee06}. By matching them at $x=0$ and $x=L$ for an electron incident from the normal contact we obtain the reflection amplitudes $r_{ee}^{(n)}(\varepsilon)$ and $r_{he}^{(n)}(\varepsilon)$. Matching for an incident hole gives the amplitudes $r_{hh}^{(n)}(\varepsilon)$ and $r_{eh}^{(n)}(\varepsilon)$. Together these four reflection amplitudes determine the scattering matrix 
\begin{equation}
S^{(n)}(\varepsilon)=
\begin{pmatrix}
r_{ee}^{(n)}(\varepsilon) & r_{he}^{(n)}(\varepsilon) \\
r_{eh}^{(n)}(\varepsilon) & r_{hh}^{(n)}(\varepsilon)
\end{pmatrix}.\label{Sdef}
\end{equation}

The result of this calculation is
\begin{subequations}
\label{Sresult}
\begin{eqnarray}
r_{eh}&=&r_{he}=\frac{1}{X}e^{i \beta}\cos\alpha_e\cos\alpha_h,\label{Sresulta}\\
r_{ee}&=&Y_e/X,\;\;r_{hh}=Y_h/X,\label{Sresultb}\\
X&=&e^{2 i \beta}(\cos\alpha_e\cos\theta_{e}-i\sin\theta_{e})(\cos\alpha_h\cos\theta_{h}-i\sin\theta_{h})\nonumber\\
&&\mbox{}-\sin\alpha_e\sin\theta_{e}\sin\alpha_h\sin\theta_{h},\label{Sresultc}\\
Y_e&=&e^{2 i \beta}\sin\alpha_e\sin\theta_{e}(\cos\alpha_h\cos\theta_{h}-i\sin\theta_{h})\nonumber\\
&&\mbox{}-\sin\alpha_h\sin\theta_{h}(\cos\alpha_e\cos\theta_{e}+i\sin\theta_{e}),\label{Sresultd}\\
Y_h&=&e^{2 i \beta}\sin\alpha_h\sin\theta_{h}(\cos\alpha_e\cos\theta_{e}-i\sin\theta_{e})\nonumber\\
&&\mbox{}-\sin\alpha_e\sin\theta_{e}(\cos\alpha_h\cos\theta_{h}+i\sin\theta_{h}).\label{Sresulte}
\end{eqnarray}
\end{subequations}
We have defined the angles
\begin{subequations}
\label{anglesdef}
\begin{eqnarray}
&&\alpha_e=\arcsin\left(\frac{\hbar vq_{n}}{\varepsilon+\mu}\right),\;\;
\alpha_h=\arcsin\left(\frac{\hbar vq_{n}}{\varepsilon-\mu}\right),\label{anglesdefa}\\
&&\theta_e=\frac{L(\varepsilon+\mu)}{\hbar v}\cos\alpha_e,\;\;
\theta_h=\frac{L(\varepsilon-\mu)}{\hbar v}\cos\alpha_h,\label{anglesdefb}\\
&&\beta=\arccos(\varepsilon/\Delta_{0}).\label{anglesdefc}
\end{eqnarray}
\end{subequations}
One can verify that $S^{(n)}(\varepsilon)$ is a unitary matrix for $\varepsilon<\Delta_{0}$.

\section{Conductance}
\label{conductance}

The linear response conductance $G_{\rm NS}=\lim_{V\rightarrow 0}\partial I/\partial V$ is given by the Blonder-Tinkham-Klapwijk formula\cite{Blo82}
\begin{eqnarray}
G_{\rm NS}&=&g_{0}\sum_{n}\bigl[1-|r_{ee}^{(n)}(0)|^{2}+|r_{he}^{(n)}(0)|^{2}\bigr]\nonumber\\
&=&2g_{0}\sum_{n}|r_{he}^{(n)}(0)|^{2}.\label{Gdef}
\end{eqnarray}
(In the second equality we used the unitarity of the scattering matrix.) The conductance quantum is $g_{0}=4e^{2}/h$, where the factor of 4 accounts for the twofold spin and valley degeneracies in graphene. The sum over modes runs from $-\mu_{\infty}W/hv$ to $+\mu_{\infty}W/hv$. In the regime $\mu_{\infty}\rightarrow\infty$, $W/L\rightarrow\infty$ of interest here, the sum over modes may be replaced by an integration over the transverse wave vector $q$,
\begin{equation}
G_{\rm NS}=2g_{0}\int_{-\infty}^{\infty}R_{A}(q)\,\frac{W}{2\pi}\,dq.\label{Gintdef}
\end{equation}
with $R_{A}(q_{n})=|r_{he}^{(n)}(0)|^{2}$ the probability for Andreev reflection at the Fermi level.

\begin{figure}[tb]
\centerline{\resizebox{0.9\linewidth}{!}{\includegraphics{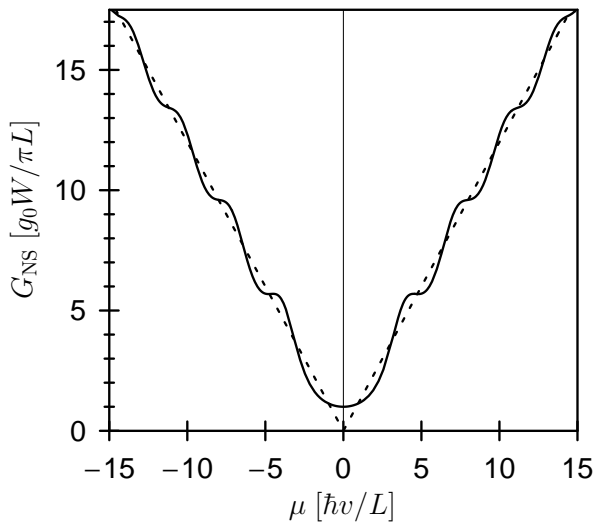}}}
\caption{\label{EV_0}
Conductance of the NS junction versus Fermi energy, calculated from Eqs.\ (\ref{Gintdef}) and (\ref{RAresult}). The dotted lines are the asymptotes (\ref{Gasymp}).}
\end{figure}

From Eq.\ (\ref{Sresult}) we obtain the expression
\begin{equation}
R_{A}(q)=\frac{k^{4}}{\bigl[(\mu/\hbar v)^{2}-q^{2}\cos(2kL)\bigr]^{2}},\label{RAresult}
\end{equation}
with $k=\sqrt{(\mu/\hbar v)^{2}-q^{2}}$ the longitudinal wave vector. For $|q|>|\mu|/\hbar v$ the longitudinal wave vector is imaginary, corresponding to an evanescent mode. Substitution of Eq.\ (\ref{RAresult}) into Eq.\ (\ref{Gintdef}) gives the conductance versus Fermi energy plotted in Fig.\ \ref{EV_0}. The asymptotes are
\begin{equation}
G_{\rm NS}=\left\{\begin{array}{ll}
\pi^{-1}g_{0}W/L&{\rm for}\;\;|\mu|\ll\hbar v/L,\\
0.38\,g_{0}|\mu| W/\hbar v&{\rm for}\;\;|\mu|\gg\hbar v/L.
\end{array}\right.\label{Gasymp}
\end{equation}

\section{Comparison with a normal junction}
\label{compare}

In Refs.\ \onlinecite{Kat06} and \onlinecite{Two06} the geometry of Fig.\ \ref{fig_NGS} was studied in the case that both contact regions are in the normal state. The transmission probability $T$ from one contact to the other in that case was found to be given by\cite{Two06}
\begin{equation}
T(q)=\frac{k^{2}}{k^{2}\cos^{2}(kL)+(\mu/\hbar v)^{2}\sin^{2}(kL)}.\label{Tresult}
\end{equation}
As before, $q$ is the transverse wave vector and $k=\sqrt{(\mu/\hbar v)^{2}-q^{2}}$ the longitudinal wave vector. In the limit $\mu\rightarrow 0$ of undoped graphene we have $T=1/\cosh^{2}(qL)$, while $R_{A}=1/\cosh^{2}(2qL)$. The effective doubling of the length in the case of a superconducting contact is consistent with the physical picture described in the introduction.

For arbitrary $\mu$, we find by comparing Eqs.\ (\ref{RAresult}) and (\ref{Tresult}) that the probability for Andreev reflection $R_{A}$ is related to the normal state transmission probability $T$ by
\begin{equation}
R_{A}=\frac{T^{2}}{(2-T)^{2}}.\label{RTrelation}
\end{equation}
This relation, derived here from the Dirac-Bogoliubov-De Gennes equation, is the same relation as for the usual (nonrelativistic) Bogoliubov-De Gennes equation.\cite{Bee92} In the Appendix we show that this relation is not restricted to ballistic NS junctions, but can be generalized to arbitrary disorder in the normal region. 

The conductance
\begin{equation}
G_{\rm N}=g_{0}\int_{-\infty}^{\infty}T(q)\,\frac{W}{2\pi}\,dq\label{GNdef}
\end{equation}
in the normal case has asymptotes\cite{Kat06,Two06}
\begin{equation}
G_{\rm N}=\left\{\begin{array}{ll}
\pi^{-1}g_{0}W/L&{\rm for}\;\;|\mu|\ll\hbar v/L,\\
\frac{1}{4}\,g_{0}|\mu| W/\hbar v&{\rm for}\;\;|\mu|\gg\hbar v/L.
\end{array}\right.\label{GNasymp}
\end{equation}
 Comparison with the asymptotes (\ref{Gasymp}) shows that the conductance is enhanced at large $|\mu|$ when one of the contacts becomes superconducting, as expected for a ballistic junction. The enhancement by a factor $1.5$ is below the factor of two enhancement expected for an ideal NS interface, because of the mismatch of Fermi wave lengths. For $\mu\rightarrow 0$ the enhancement disappears and $G_{\rm NS}$ becomes precisely equal to $G_{\rm N}$, as a manifestation of pseudo-diffusive conduction.

Since the first experiments on induced superconductivity and Andreev reflection in graphene have now been reported,\cite{Hee06,Gei06} our theoretical predictions may well be tested in the near future.

\acknowledgments
This research was supported by the Dutch Science Foundation NWO/FOM.

\appendix
\section{General relation between the conductance of an NS junction in graphene and its normal-state value}

In Ref.\ \onlinecite{Bee92} a general relation was derived between the conductance $G_{\rm NS}$ of an NS junction and its normal-state value $G_{\rm N}$, on the basis of the Bogoliubov-De Gennes (BdG) equation.\cite{deG66} Here we show that the same relation holds also for an NS junction in graphene, described by the Dirac-Bogoliubov-De Gennes (DBdG) equation.\cite{Bee06}

The relation of Ref.\ \onlinecite{Bee92} is expressed by the pair of equations
\begin{equation}
G_{\rm NS}=2g_{0}\sum_{n}\frac{T_{n}^{2}}{(2-T_{n})^{2}},\;\;G_{\rm N}=g_{0}\sum_{n}T_{n},\label{GNSGNrelation}
\end{equation}
where $g_{0}=se^{2}/h$ is the conductance quantum (including an $s$-fold spin and valley degeneracy factor). The transmission eigenvalues $T_{n}$ are the eigenvalues of the matrix product $tt^{\dagger}$, with $t$ the transmission matrix at the Fermi level when both contacts are in the normal state. The region in between the two contacts may be arbitrarily disordered, so $t$ is no longer a diagonal matrix [as in the ballistic case considered in the text, leading to Eq.\ (\ref{RTrelation})]. We also allow for intervalley scattering, in which case $g_{0}$ includes only the spin degeneracy ($s=2$).

The relation (\ref{GNSGNrelation}) relies on a separation of length scales on the superconducting side of the NS interface: The first length scale $\lambda'_{F}=hv/\mu_{\infty}$ determines the distance over which the step (\ref{Udef}) in the electrostatic potential $U(x)$ affects the wave functions in S. The second length scale $\xi=\hbar v/\Delta_{0}$ determines the distance over which the step (\ref{Deltadef}) in the pair potential $\Delta(x)$ affects the wave functions in S. Since $\lambda'_{F}\ll\xi$, we may treat the scattering by $U$ and $\Delta$ independently of each other.

Setting first $\Delta\equiv 0$, we consider the normal-state scattering problem, including the electrostatic potential $U$ and any disorder in the region $x>0$. This determines the scattering matrix $S_{0}(\varepsilon)$ for electrons and its counterpart $S_{0}^{\ast}(-\varepsilon)$ for holes. There is no mixing of electrons and holes for $\Delta\equiv 0$, so the full normal-state scattering matrix has the form
\begin{equation}
{\cal S}_{\rm N}(\varepsilon)=\begin{pmatrix}
S_{0}(\varepsilon)&0\\
0&S_{0}^{\ast}(-\varepsilon)
\end{pmatrix}.\label{SNdef}
\end{equation}
The basis of hole states $\Psi_{h}$ is chosen such that it is the time reversed of the basis of electron states $\Psi_{e}$: $\Psi_{h}(\varepsilon)=(\sigma_{z}\otimes\tau_{x})\Psi_{e}^{\ast}(-\varepsilon)$, with Pauli matrices $\sigma_{i}$ and $\tau_{i}$ acting, respectively, on the isospin and valley degree of freedom.\cite{Bee06} The normal-state reflection and transmission matrices are submatrices of $S_{0}$,
\begin{equation}
S_{0}=\begin{pmatrix}
r&t'\\
t&r'
\end{pmatrix}.\label{S0def}
\end{equation}
Since we assume zero magnetic field, $S_{0}$ is symmetric as well as unitary.

Setting then $U\equiv 0$, we consider the Andreev reflection by the superconductor at $x<0$ (in the absence of any disorder for $x>0$). The scattering matrix for Andreev reflection at subgap energies $\varepsilon<\Delta_{0}$ is given by
\begin{equation}
{\cal S}_{A}(\varepsilon)=\zeta(\varepsilon)\begin{pmatrix}
0&1\\
1&0
\end{pmatrix},\;\;\zeta(\varepsilon)=\exp[-i\arccos(\varepsilon/\Delta_{0})].\label{SAdef}
\end{equation}
This has the same form in the BdG and DBdG equation.

The total reflection matrix $S$ is constructed from $S_{\rm N}$ and $S_{A}$ in the same way as in Ref.\ \onlinecite{Bee92}. We need the submatrix
\begin{equation}
r_{he}(\varepsilon)=\zeta(\varepsilon)t^{\ast}(-\varepsilon)[1-\zeta^{2}(\varepsilon)r(\varepsilon)r^{\ast}(-\varepsilon)]^{-1}t'(\varepsilon).\label{rhe}
\end{equation}
Setting $\varepsilon=0$ and using the symmetry and unitarity of $S_{0}$, the conductances
\begin{equation}
G_{\rm NS}=2g_{0}{\rm Tr}\,r_{he}^{\vphantom{\dagger}}(0)r_{he}^{\dagger}(0),\;\;G_{\rm N}=g_{0}{\rm Tr}\,t(0)t^{\dagger}(0)\label{GNSGNdef}
\end{equation}
take the form of Eq.\ (\ref{GNSGNrelation}).

\end{document}